\pdfoutput=1
\documentclass[10pt,a4paper,twoside]{iopart}
\usepackage{iopams}
\expandafter\let\csname equation*\endcsname\relax
\expandafter\let\csname endequation*\endcsname\relax
\usepackage{amsmath,amsfonts,amssymb}
\usepackage{graphicx}
\usepackage[T1]{fontenc}
\usepackage[utf8]{inputenc}
\usepackage{microtype}
\usepackage[numbers,square,sort&compress]{natbib}
\usepackage[textsize=footnotesize]{todonotes}
\usepackage{soul}
\usepackage{hyperref}
\usepackage[noetal,notodos]{CMImacros}

\newcommand{\Ip}{\ensuremath{\text{I}^{+}}\xspace}
\newcommand{\Ipp}{\ensuremath{\text{I}^{+2}}\xspace}
\newcommand{\Ippp}{\ensuremath{\text{I}^{+3}}\xspace}
\newcommand{\Ipn}{\ensuremath{\text{I}^{+n}}\xspace}

\begin{document}
\setcounter{page}{1}%
\title{Strongly aligned gas-phase molecules at Free-Electron Lasers}%
\author{%
   \mbox{Thomas Kierspel},$\!^{1,2}$ %
   \mbox{Joss Wiese},$\!^{1}$ %
   \mbox{Terry Mullins},$\!^{1}$ %
   \mbox{Joseph Robinson},$\!^{3}$ %
   \mbox{Andy Aquila},$\!^{3}$ %
   \mbox{Anton Barty},$\!^{1}$ %
   \mbox{Richard Bean},$\!^{1,4}$ %
   \mbox{Rebecca Boll},$\!^{5}$ %
   \mbox{Sébastien Boutet},$\!^{3}$ %
   \mbox{Philip Bucksbaum},$\!^{3,6,7}$ %
   \mbox{Henry N. Chapman},$\!^{1,2,8}$ %
   \mbox{Lauge Christensen},$\!^{9}$ %
   \mbox{Alan Fry},$\!^{3,6}$ %
   \mbox{Mark Hunter},$\!^{3}$ %
   \mbox{Jason E.\ Koglin},$\!^{3}$ %
   \mbox{Mengning Liang},$\!^{3}$ %
   \mbox{Valerio Mariani},$\!^{1}$ %
   \mbox{Andrew Morgan},$\!^{1}$ %
   \mbox{Adi Natan},$\!^{6}$ %
   \mbox{Vladimir Petrovic},$\!^{6}$ %
   \mbox{Daniel Rolles},$\!^{5,10}$ %
   \mbox{Artem Rudenko},$\!^{10}$ %
   \mbox{Kirsten Schnorr},$\!^{11}$ %
   \mbox{Henrik Stapelfeldt},$\!^{9}$ %
   \mbox{Stephan Stern},$\!^{1}$ %
   \mbox{Jan Th{\o}gersen},$\!^{9}$ %
   \mbox{Chun Hong Yoon},$\!^{1,4}$ %
   \mbox{Fenglin Wang},$\!^{1,6}$ %
   \mbox{Sebastian Trippel},$\!^{1}$ %
   and \mbox{Jochen~Küpper}$\!^{1,2,8}$%
}%
\ead{\href{mailto:jochen.kuepper@cfel.de}{jochen.kuepper@cfel.de};
   website:~\url{http://www.controlled-molecule-imaging.org}}%
\address{\noindent%
   \begin{itemize}
   \item[$^1$~] Center for Free-Electron Laser Science, DESY, 22607 Hamburg, Germany
   \item[$^2$~] Center for Ultrafast Imaging, University of Hamburg, 22761 Hamburg, Germany
   \item[$^3$~] LCLS, SLAC National Accelerator Laboratory, Menlo Park, CA, 94025, USA
   \item[$^4$~] European X-ray Free Electron Laser (XFEL) GmbH, 22761 Hamburg, Germany
   \item[$^5$~] Deutsches Elektronen-Synchrotron (DESY), 22607 Hamburg, Germany
   \item[$^6$~] SLAC National Accelerator Laboratory, PULSE Institute, Stanford, CA, 94305, USA
   \item[$^7$~] Department of Physics, Stanford University, Stanford, CA 94305, USA
   \item[$^8$~] Department of Physics, University of Hamburg, 22761 Hamburg, Germany
   \item[$^9$~] Department of Chemistry, Aarhus University, 8000 Aarhus C, Denmark
   \item[$^{10}$~] Department of Physics, Kansas State University, Manhatten, KS, 66506, USA
   \item[$^{11}$~] Max Planck Institute for Nuclear Physics, 69117 Heidelberg, Germany
   \end{itemize}
}%
\begin{abstract}%
   We demonstrate a novel experimental implementation to strongly align molecules at full repetition
   rates of free-electron lasers. We utilized the available in-house laser system at the coherent
   x-ray imaging beamline at the Linac Coherent Light Source. Chirped laser pulses, \ie, the direct
   output from the regenerative amplifier of the Ti:Sa chirped pulse amplification laser system,
   were used to strongly align 2,5-diiodothiophene molecules in a molecular beam. The alignment
   laser pulses had pulse energies of a few \mJ and a pulse duration of 94~ps. A degree of alignment
   of \cost~=~0.85 was measured, limited by the intrinsic temperature of the molecular beam rather
   than by the available laser system. With the general availability of synchronized
   chirped-pulse-amplified near-infrared laser systems at short-wavelength laser facilities, our
   approach allows for the universal preparation of molecules tightly fixed in space for experiments
   with x-ray pulses.
\end{abstract}
\enlargethispage{15mm}
\begin{indented}
\item[]\today \\
\item[]\noindent{}Submitted to: \JPB%
\item[]\emph{Keywords}: FEL, Gas Phase, Alignment
\end{indented}
\noindent\ioptwocol
\section{Introduction}%
\label{sec:introduction}%
Imaging the structural dynamics of gas-phase molecules using approaches such as photoelectron
imaging in the molecular frame~\cite{Meckel:Science320:1478, Holmegaard:NatPhys6:428,
   Boll:PRA88:061402}, high-harmonic-generation spectroscopy~\cite{Itatani:Nature432:867,
   Woerner:Nature466:604}, laser-induced electron diffraction~\cite{Zuo:CPL159:313,
   Blaga:Nature483:194}, electron diffraction~\cite{Hensley:PRL109:133202, Sciaini:RPP74:096101}, or
x-ray diffraction~\cite{Kuepper:PRL112:083002, Barty:ARPC64:415, Stern:JPB:inprep} is much improved
or is only possible by fixing the molecules in space~\cite{Stapelfeldt:RMP75:543, Barty:ARPC64:415,
   Stern:JPB:inprep, Filsinger:PCCP13:2076, Hensley:PRL109:133202, Kuepper:PRL112:083002}. Various
approaches to align or orient molecules have been demonstrated, ranging from
state-selection~\cite{Reuss:StateSelection} over brute-force orientation~\cite{Block:PRL68:1303} to
laser alignment~\cite{Stapelfeldt:RMP75:543}. Generally, laser alignment has been implemented in two
different regimes, namely, using short laser pulses to ``impulsively'' create coherent rotational
wavepackets that generate alignment some time after the laser pulse~\cite{Normand:JPB25:497,
   RoscaPruna:PRL87:153902} or using long laser pulses to ``adiabatically'' create pendular states
that are strongly aligned during the laser pulse~\cite{Sakai:JCP110:10235}. In the former approach,
pulses from standard commercial Ti:Sapphire (TSL) laser systems can be utilized at repetition rates
up to several kilohertz, whereas in the latter approach, traditionally, injection seeded Nd:YAG
lasers were employed, with hundreds of millijoules of pulse energy, limited to a few 10~Hz. Through
the addition of weak dc electric fields, strong orientation can be achieved in both
scenarios~\cite{Holmegaard:PRL102:023001, Ghafur:NatPhys5:289, Filsinger:JCP131:064309}.
Three-Dimensional (3D) alignment or orientation was also demonstrated~\cite{Larsen:PRL85:2470,
   Lee:PRL97:173001, Ren:PRL112:173602, Tanji:PRA72:063401, Holmegaard:PRL102:023001,
   Filsinger:JCP131:064309}. It was shown that the degree of alignment necessary for molecular-frame
diffractive imaging is very high~\cite{Pabst:PRA81:043425, Filsinger:PCCP13:2076, Barty:ARPC64:415}
and, for complex molecules, could not be achieved with impulsive alignment approaches. Generally,
the degree of alignment can be improved through higher laser-field intensities, limited by the onset
of ionization, or colder samples~\cite{Kumarappan:JCP125:194309}, possibly achieved through state
selection \cite{Holmegaard:PRL102:023001, Ghafur:NatPhys5:289, Filsinger:JCP131:064309}.

The intermediate-pulse-duration regime, for which the alignment laser pulse duration is shorter than
the lowest rotational period of the molecule, but cannot be considered impulsive anymore, had
previously been analyzed theoretically~\cite{Ortigoso:JCP110:3870, Torres:PRA72:023420,
   Owschimikow:PCCP13:8671}. While the laser pulse is still on, the molecules act similar to the
adiabatic alignment case, whereas after the pulse periodic revivals can be
observed~\cite{Ortigoso:JCP110:3870}. Recently, this mesobatic regime was explored experimentally
and it was realized that degrees of alignment comparable to the adiabatic limit can be achieved with
much shorter and weaker laser pulses, \eg, from commercial chirped-pulse-amplified (CPA) TSL
systems~\cite{Trippel:MP111:1738, Trippel:PRA89:051401R}. While the nonadiabatic interaction creates
interesting dynamic phenomena such as the observation of pendular
motion~\cite{Trippel:PRA89:051401R}, this approach does provide strongly aligned samples of complex
molecules~\cite{Trippel:MP111:1738}. Moreover, it requires only moderately long pulses on the order
of $100$~ps with moderate pulse energies, \ie, a few mJ, which are available as output from the
standard laser systems available at x-ray laser facilities.

Here, we demonstrate the successful implementation of intermediate-pulse-duration regime, strong
alignment of the prototypical asymmetric rotor molecule 2,5-diiodothiophene
($\text{C}_4\text{H}_2\text{I}_2\text{S}$) in a molecular beam at the Coherent X-ray Imaging
(CXI)~\cite{Boutet:NJP12:035024, Liang:JSR22:514} endstation of the Linac Coherent Light Source
(LCLS) at the SLAC National Accelerator Laboratory (SLAC). Previous experiments with fixed-in-space
molecules at free-electron lasers (FELs) have exploited both, impulsive and adiabatic alignment, for
instance, in ion imaging~\cite{Johnsson:JPB42:134017, Glownia:OE18:17620}, electron imaging~
\cite{Cryan:PRL105:083004, Boll:PRA88:061402, Rolles:JPB47:124035, Boll:FD17171:57}, or x-ray
diffraction experiments~\cite{Kuepper:PRL112:083002, Stern:FD171:393}. Our approach allows one to
combine the strong alignment achieved in adiabatic alignment approaches~\cite{Boll:PRA88:061402,
   Kuepper:PRL112:083002, Stern:FD171:393, Boll:FD17171:57, Rolles:JPB47:124035} with the
utilization of x-ray pulses at the full repetition rate of the LCLS~\cite{Glownia:OE18:17620,
   Cryan:PRL105:083004}. This allowed to perform experiments at 120~Hz, comparable to previously
demonstrated intermediate regime alignment at 1~kHz in table-top
experiments~\cite{Trippel:MP111:1738}. For comparison, we also provide measurements for the degree
of alignment from experiments conducted in our laboratory at the Center for Free-Electron Laser
Science (CFEL)~\cite{Trippel:MP111:1738} during the beam time preparation.


\section{Experimental Setup}
\label{sec:experiment}
\begin{figure}[t]
   \centering
   \includegraphics[width=\linewidth]{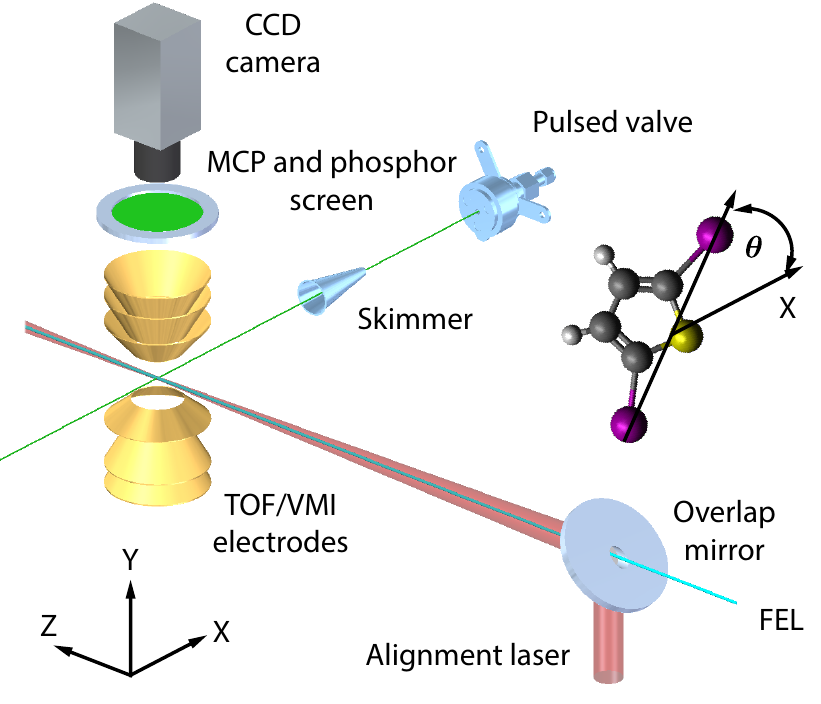}
   \caption{Sketch of the experimental setup showing the pulsed molecular beam valve, the laser beam
      path of the alignment laser (red) and the free-electron laser (cyan), and a holey mirror to
      collinearly overlap both beams. The detection system consists of an ion-imaging setup for
      time-of-flight measurements and velocity-map imaging onto a micro-channel plate, phosphor
      screen, and a CCD camera. In the inset the definition of $\theta$ is depicted, it is the angle
      between the laser polarization, \ie, the $X$ axis, and the most polarizable axis of the
      molecule, \ie, the iodine-iodine axis.}
   \label{fig:setup}
\end{figure}
\autoref{fig:setup} shows a sketch of the experimental setup at LCLS. A trace of 2,5-diiodothiophene
was coexpanded in 80~bar helium into the vacuum by a pulsed valve~\cite{Even:JCP112:8068}, operated
at \celsius{75}, and at a repetition rate of 120~Hz synchronized to the LCLS. The molecular beam was
skimmed 8~cm downstream the nozzle using a 3-mm-diameter skimmer. This resulted in a 5.2~mm
molecular beam (full width at half maximum, FWHM) in the interaction zone.

Near-infrared (IR) laser pulses were produced in a CPA TSL system (Coherent) with a central
wavelength of 800~nm. The laser system is operated at and synchronized to the repetition rate of
LCLS. The chirped pulses out of the regenerative amplifier were split using a beam splitter with a
reflectivity of 90~\%. The reflected pulses, which were used to align the
molecules~\cite{Trippel:MP111:1738,Trippel:PRA89:051401R}, had a pulse energy of 3.3~\mJ and a pulse
duration of 94~ps (FWHM). The transmitted pulses were compressed using a standard grating-based
compressor to a pulse duration of 55~fs (FWHM) at a pulse energy of 250~\textmu{}J. The compressed
pulses multiply ionize the molecules through strong-field ionization, resulting in Coulomb
explosion; in all experiments these compressed pulses were linearly polarized, with the polarization
perpendicular to the long axis of polarization of the alignment laser pulses (\emph{vide infra}).
Both arms were spatially and temporally recombined before entering the vacuum system. This, in
combination with ion imaging (\emph{vide infra}), allows to characterize and optimize the degree of
alignment without the FEL beam.

The short pulses were also utilized to measure the pulse duration of the alignment laser pulses
through spectral interference. The frequency components of the chirped pulse arrive sequentially,
whereas all frequencies in the compressed pulse are temporally overlapped. At finite spectrometer
resolution, interference in the combined spectrum is only visible at those wavelength components of
the alignment pulse that are temporally overlapped, or nearly overlapped, with the compressed pulse.
The strongest interference is visible at the wavelength that temporally coincides with the
compressed pulse. A delay line was used to change the relative timing between the chirped alignment
and the compressed pulses. Thus, monitoring the spectrum in dependence of the delay between these
two pulses provided information on the pulse duration of the chirped pulse.

To reach sufficiently high intensities to align the molecules, a telescope was used to expand both
beams by a factor of three before being focused into the chamber with a 1-m-focal-length lens. A
holey mirror was used to spatially overlap the TSL pulses with the FEL beam path, losing 20~\% of
the near-infrared laser power. The hole had a diameter of 2~mm and the FEL beam was guided through
the hole to achieve a collinear propagation of both beams. The alignment laser focus size was 45~\um
(FWHM) leading to an estimated peak intensity at the focus on the order of
1$\cdot$1$0^{12}$~$\text{W}/\text{cm}^2$. The FEL beam size was estimated to 12~\um in the
horizontal and 3~\um in the vertical axis. The different beam diameters made the setup more stable
against beam pointing imperfections and ensured that only well aligned molecules were probed with
the FEL. The alignment laser pulse and the FEL/TSL pulses were spatially and temporally overlapped
in order to maximize the degree of alignment; the exact timing was not
critical~\cite{Trippel:MP111:1738}.

The alignment laser beam was elliptically polarized with an aspect ratio of 3:1 in order to obtain 3D
alignment. The major axis was parallel to the micro-channel plate (MCP) and phosphor screen surface.
That way the most polarizable axis of 2,5-diiodothiophene, \ie, the iodine--iodine axis, was
confined to the $X$ axis as can be seen in \autoref{fig:setup}. The degree of alignment was probed
using x-ray pulses with a pulse duration of approximately 70 fs (FWHM), a photon energy of 9.5~keV,
and a pulse energy of approximately 0.64~\mJ ($4.2\cdot10^{11}$ photons, beam line transmission
80~\%, focussing mirror transmission 40~\%) in the interaction zone. X-ray absorption is largely
localized at one of the iodine atoms, followed by Auger cascades and intramolecular charge
redistribution leading to Coulomb explosion of the molecule. The ions were accelerated by the
electric field of the spectrometer (\autoref{fig:setup}) towards a position sensitive detector
consisting of a MCP and a phosphor screen (Photonis). A CCD camera (Adimec Opal), operating at the
FEL repetition rate, recorded single-shot ion distributions visible as light spots on the phosphor
screen. A real time online monitoring software, implementing a peak
finding~\cite{Barty:JAPPCRY:2014} and centroiding~\cite{Kella:NIMA329:440, Chang:RSI69:1665}
algorithm, was used to identify and localize single ion hits on the detector. By measuring the
transverse velocity distribution of the iodine cations, the degree of alignment is evaluated by
determining \cost\footnote{The two-dimensional degree of alignment is defined as
   \mbox{$\cost=\int_{0}^{\pi}\int_{v_min}^{v_{max}}\!\cos^2\!\left(\theta_\text{2D}\right)f(\thetatwoD,v_{2D})\,dv_{2D}d\thetatwoD$},
   where $f(\thetatwoD,v_{2D})$ is the normalized 2D projection of the probability density.}, where
$\theta$ is defined as indicated in \autoref{fig:setup} and \thetatwoD is defined as indicated in
\autoref{fig:VMI}.

The detection system allowed to switch between a time-of-flight (TOF)~\cite{Wiley:RSI26:1150} mode
to determine the various fragment masses and a velocity map imaging (VMI)~\cite{Eppink:RSI68:3477}
mode to determine the ion angular distribution. In the VMI mode, a fast high-voltage switch (Behlke)
was used to gate the phosphor screen on a single iodine charge state determined from the TOF mass
spectrum (MS).

The experimental details for the comparison measurement performed at the CFEL are described in
reference~\citealp{Trippel:MP111:1738}. In short, the molecular beam was triply skimmed before it
entered a VMI spectrometer where it was crossed by the alignment and probe laser pulses. The
distance from the valve to the interaction zone was approximately a factor of 3.5 longer than at the
LCLS. The sample reservoir was heated to \celsius{40} and a stagnation pressure of 90~bar helium was
applied. The alignment and probe laser pulses were provided by splitting the output of an amplified
femtosecond TSL system into two parts. The alignment pulses were stretched, or, more accurately,
overcompressed. This resulted in negatively chirped pulses with a duration of 485~ps (FWHM) and
focused to a beam waist of 36~$\mu$m times 39~$\mu$m (FWHM). At a maximum pulse energy of 7.4 mJ a
focus peak intensity of 9.6$\cdot$1$0^{11}$~$\text{W}/\text{cm}^2$ can be estimated. The temporal
profile of the alignment pulse had a trapezoid-like shape, \ie, fast rising time (100~ps), longer
plateau (400~ps), and fast falling time (100~ps). The alignment laser was elliptically polarized
with an aspect ratio of 3:1. The probe pulses were compressed to 30~fs and focused to
$27\times28~\um^2$ (FWHM).

\section{Experimental Results}
\subsection{Results from experiments at LCLS}
\label{sec:results_LCLS}
\begin{figure}
   \centering
   \includegraphics[width=\linewidth]{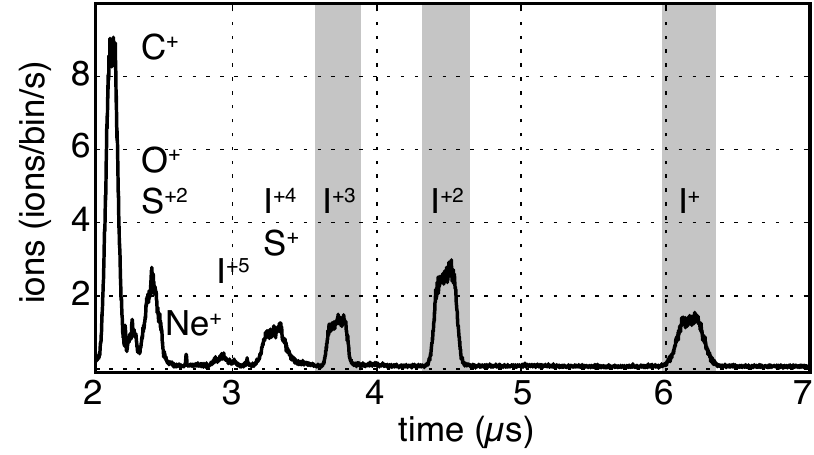}
   \caption{Time-of-flight mass spectrum of the ionic fragments produced in the Coulomb explosion of
      2,5-diiodothiophene following absorption of a 9.5~keV x-ray photon. The gray areas indicate
      the gating windows for the VMI spectrometer used to determine the angular distribution for each
      fragment in \autoref{fig:VMI}.}
   \label{fig:TOF}
\end{figure}
Following the absorption of a 9.5~keV x-ray photon the 2,5-diiodothiophene molecules charge up and
fragment through Coulomb explosion. \autoref{fig:TOF} shows a time-of-flight (TOF) mass spectrum
(MS) build up as a histogram of events from several thousand single-shot mass spectra. Clearly
separated peaks for singly to triply charged iodine ions are visible, while the peaks for higher
charge states of iodine ions overlap with other fragments. Based on this TOF spectrum, we have
recorded angular distributions of the \Ip, \Ipp, and \Ippp fragments with the VMI spectrometer by
gating the detector for the appropriate arrival times, as indicated by the gray areas in
\autoref{fig:TOF}.

\begin{figure}[t]
   \centering
   \includegraphics[width=\linewidth]{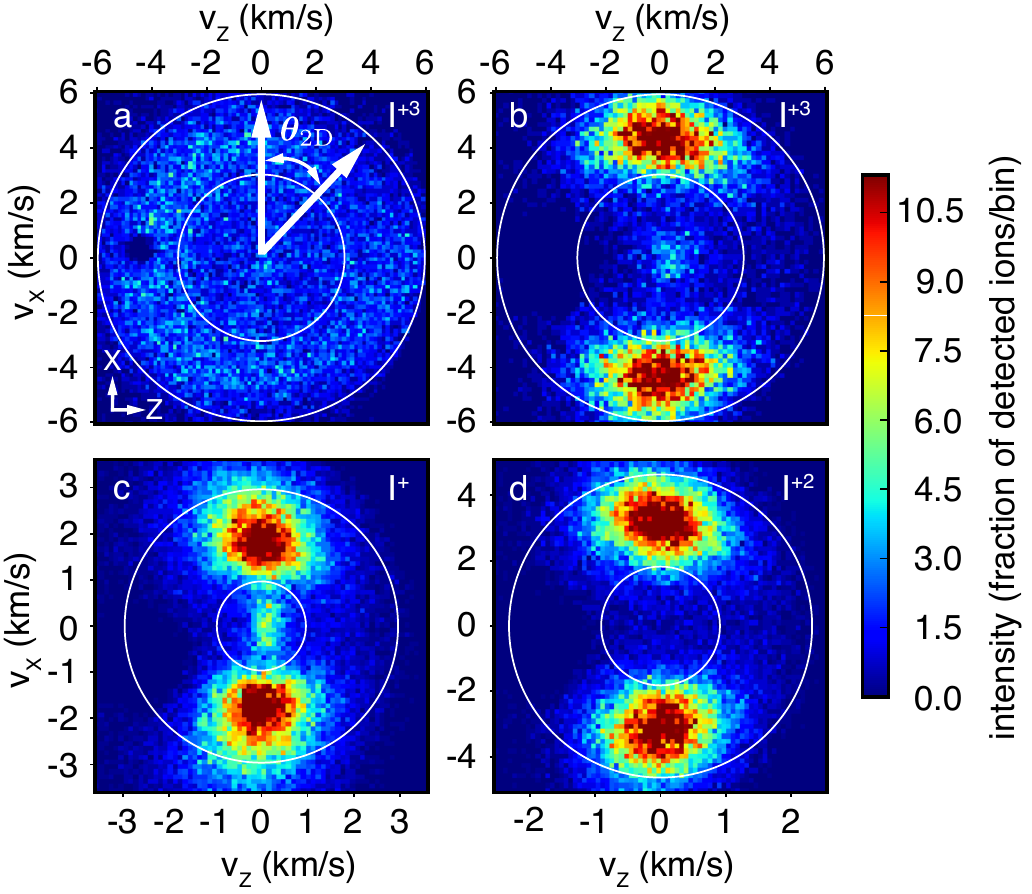}
   \caption{VMI images for (a) isotropic and (b--d) aligned iodine cations. The area between the
      inner and outer circle was used to determine the degree of alignment. The laboratory framework
      and the definition of the angle \thetatwoD is shown in a). The dark spot in the left middle is
      due to a damaged area on the MCP, but it has no significant influence on our results; see text
      for details.}
   \label{fig:VMI}
\end{figure}
In \autoref{fig:VMI} a sum of several thousand individually-processed VMI images for several iodine
cation fragments of 2,5-diiodothiophene upon ionization with a 9.5~keV x-ray photon are shown. When
the alignment laser was blocked, the image, \autoref{fig:VMI}\,a, shows a practically isotropic
velocity distribution for triply charged iodine ions obtained from Coulomb explosion of randomly
oriented molecules. As expected for the isotropic sample the measured degree of alignment is
$\cost=0.50$. \autoref{fig:VMI}\,b--d show the corresponding velocity distributions for triply,
singly and doubly charged iodine fragments received from aligned molecules. The majority of detected
ions is located at two distinct spots around $\thetatwoD=0$ and $\pi$. The non-isotropic ion
fragment distribution is a clear indication that the molecules are aligned along the $X$ axis, the
major axis of the polarization ellipse of the alignment laser. The velocity distributions show
additional structure due to different fragmentation channels of 2,5-diiodothiophene upon x-ray
ionization, but only the areas between the inner and outer white circles were used to determine the
degree of alignment. This corresponds to the fastest observed ion fragment for each iodine charge
state. These fragments are a good measure of the degree of alignment since they are produced from
molecules undergoing the fastest fragmentation, thus exhibiting the most-axial recoil. We point out
that non-axial recoil would result in a smaller measured degree of alignment than actually exists
and, therefore, that the measurements provide a lower limit for the actual degree of alignment. The
resulting degrees of alignment for triply, doubly, and singly charged iodine cations are
$\cost=0.85$, $\cost=0.84$, and $\cost=0.82$, respectively. The dark spot in the left part of the
images, best visible in \autoref{fig:VMI}\,a, is due to a damaged area on the MCP. It has no
significant influence on the measured degree of alignment as the velocity distribution of the ions,
from aligned molecules, is peaked at $\thetatwoD=0$ and $\pi$. The tilt of the \Ipn ion
distributions in the VMIs is attributed to imperfect imaging due to a off-center ionization.

\subsection{Results from experiments at CFEL}
\label{sec:results_cfel}
\begin{figure}[t]
   \centering
   \includegraphics[width=\linewidth]{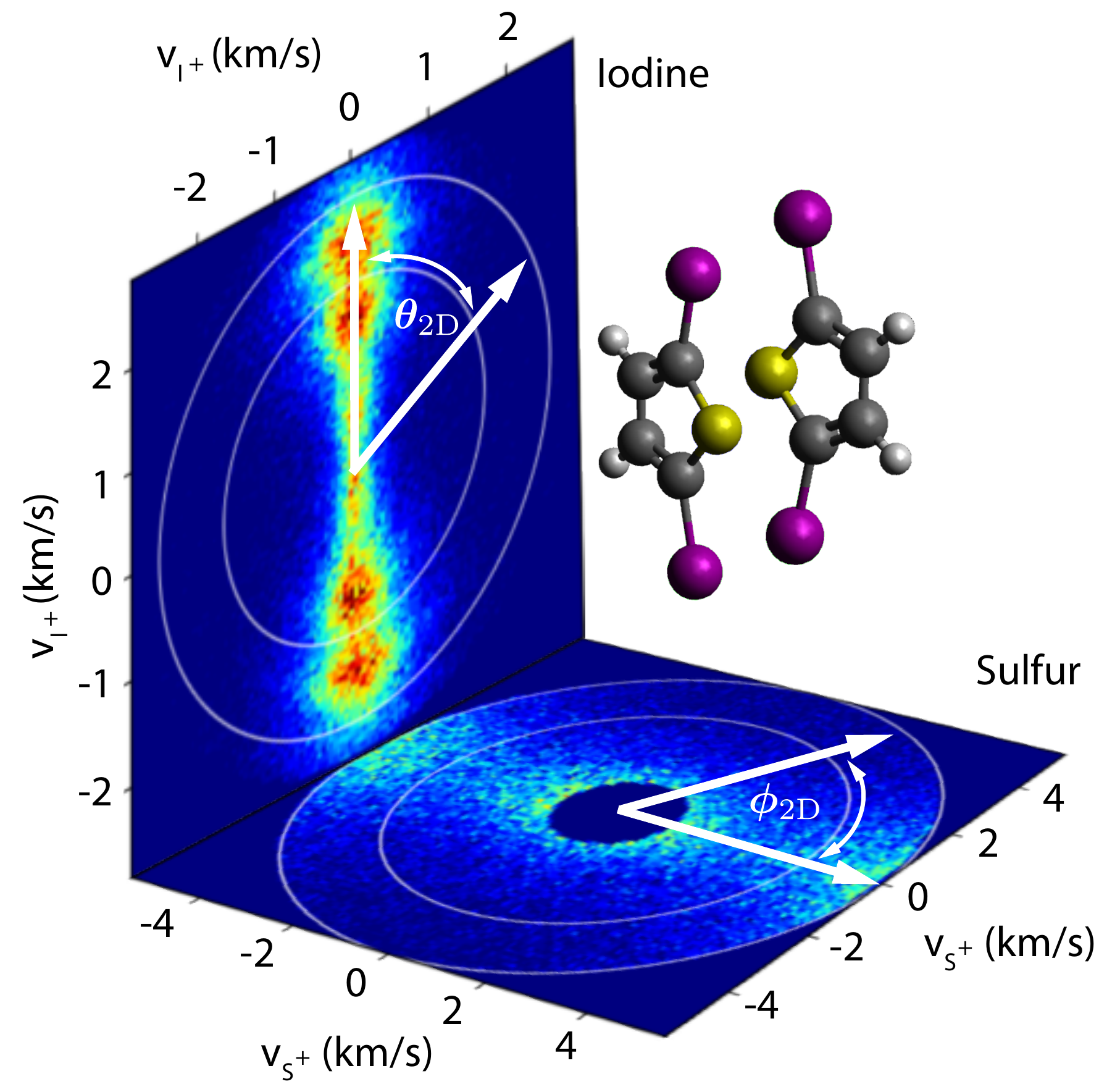}
   \caption{VMI images for singly charged iodine (side view, vertical plane) and sulfur (top view,
      horizontal plane) measured upon near-infrared strong field ionization. The two molecules
      sketch the 3D alignment. The area between the inner and outer white circles was used to
      determine the degree of alignment. Background from low-energy fragmentation channels were
      removed in the sulfur VMI. The color scale is the same as for \autoref{fig:VMI}. The white
      arrows in the images depict the angles $\thetatwoD$ and $\phitwoD$.}
   \label{fig:vmicfel}
\end{figure}
\autoref{fig:vmicfel} shows 2D momentum distributions for singly charged iodine (side view, vertical
plane) and sulfur fragments (top view, horizontal plane) upon near-infrared (NIR) strong-field
ionization (SFI) of 3D aligned molecules. The major (minor) axis of the alignment laser polarization
ellipse was set along the $X$ axis ($Y$ axis). The corresponding spatial orientation of 3D aligned
2,5-diiodothiophene is indicated by the two molecules. For singly charged iodine ions three major
Coulomb fragmentation channels are visible by the three maxima in the transverse velocity. The
velocity distribution of the fragmentation channel with the highest kinetic energy (indicated by the
white circles) was used to determine the degree of alignment as $\cost=0.94$, reflecting the very
strong alignment of the iodine-iodine axis. For the sulfur ions, two major fragmentation channels
are observed. The slow channel, located in the center of the image, is significantly stronger in
signal and has been removed to increase the contrast for the fast channel. The degree of alignment
obtained from the velocity distribution of the fast sulfur ions is $\cosp=0.79$, reflecting the
additional strong alignment of the molecular plane to the plane of polarization of the alignment
laser.

\begin{figure}
   \centering \includegraphics[width=\linewidth]{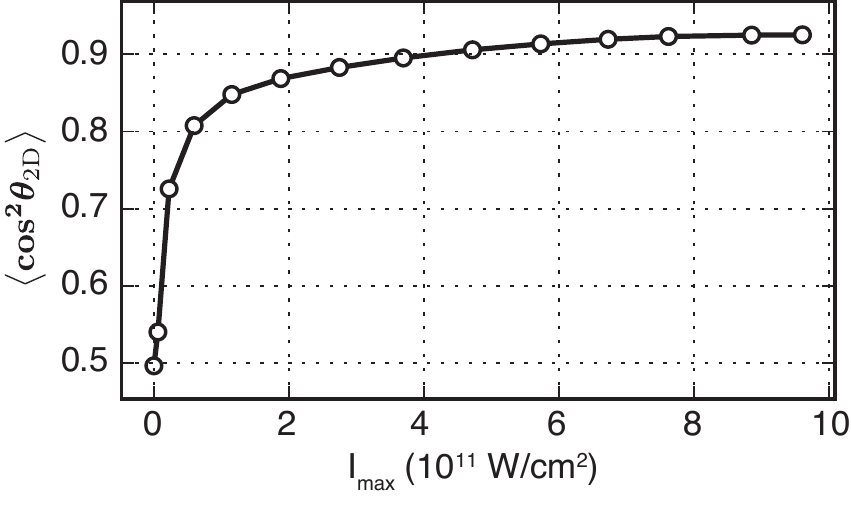}
   \caption{The degree of alignment for singly charged iodine as a function of the peak intensity of
      the alignment laser using a pulse duration of 485~ps (FWHM). This experiment was conducted at
      CFEL.}
   \label{fig:power}
\end{figure}
In \autoref{fig:power} the degree of alignment of the iodine-iodine axis is shown as a function of
the laser pulse energy, obtained from \Ip momentum images. The typical power dependence of the
degree of alignment for a cold beam is observed. The highest degree of alignment of $\cost=0.94$ is
obtained at a peak intensity of $9.6\cdot10^{11}~\text{W}/\text{cm}^2$, but above
$7\cdot10^{11}~\text{W}/\text{cm}^2$ no significant change of the degree of alignment can be seen,
indicating the saturation of the degree of alignment.

\section{Discussion}
\label{sec:discussion}
The principal aim of the experiment performed at LCLS was to record the x-ray diffraction patterns
of isolated gas phase molecules, similar to previous experiments~\cite{Kuepper:PRL112:083002,
   Stern:FD171:393}, which will be reported elsewhere. The experiments at CFEL were performed in
preparation of that beamtime. Since in the diffractive imaging experiment the amount of scattered
photons scales linearly with the density of the molecules in the molecular beam, the experimental
setup was designed such that the pulsed valve was located as close as possible to the interaction
region for a collimated, dense, cold molecular beam. A density estimation of the molecular beam can
be derived from the mass spectrum, \autoref{fig:TOF}. To obtain a lower boundary for the molecular
beam density it was assumed that all ionized molecules show two ionic iodine fragments and that the
total ion-detection efficiency is 50~\%. This led to an average of 8.6~ionized molecules per x-ray
pulse. X-ray photons mainly interact with inner shells electrons and therefore the total molecular
photo-absorption cross section is approximated by the sum of the individual atomic photo-absorption
cross sections. At a photon energy of 9.5~keV the atomic photo-absorption cross sections are
$3.86\cdot10^{-20}~\text{cm}^2$ for iodine, $3.05\cdot10^{-21}~\text{cm}^2$ for sulfur, and
$4.88\cdot10^{-23}~\text{cm}^2$ for carbon~\cite{Stern:thesis:2013, Berger:XCOM:V15}. The
interaction length is given by the FWHM of the molecular beam. This results in a molecular beam
density of $1\cdot10^{9}$~molecules~per~$\text{cm}^{-3}$.

To determine the degree of alignment it was assumed that the iodine ions recoil along the most
polarizable axis of the molecules, which is parallel to the iodine--iodine axis. This lead to an
underestimation of the degree of alignment since positive charges can also be created at different
locations in the molecule, for instance at the sulfur atom, leading to a non-axial recoil of the
iodine ions. The difference in the degree of alignment between \Ip, \Ipp, \Ippp can be explained by
the stronger Coulomb repulsion for higher charge states, resulting in faster fragmentation. This
provided a more instantaneous mapping of the molecular orientation. Thus a more accurate, less
underestimated, value for the degree of alignment is measured for higher charged states.

For the 3D alignment measurement conducted at CFEL, the difference in alignment of the two different
molecular axes ($\cost$ vs.\ $\cosp$) is due to the fact that the most polarizable axis, the
iodine-iodine axis, aligns along the long axis of the laser polarization, resulting in the strongest
interaction. The second-most polarizable axis of the molecule aligns along the short axis of laser
polarization, resulting in a somewhat weaker interaction. For the rigid molecule investigated here,
the third axis can safely be assume to be fixed with respect to the two others, and thus shows the
same degree of alignment as the secondary axis. We note that all degrees of alignment are specified
as projections on the experimental screen, as is typically done in experiments on comparably large
and complex molecules~\cite{Larsen:JCP111:7774, Holmegaard:PRL102:023001, Hansen:JCP139:234313,
   Trippel:MP111:1738}.

The difference between the measured degree of alignment at LCLS and CFEL can be understood as
follows: The major factors that affect the achievable degree of alignment are the peak intensity of
the alignment laser pulse, the rise time of the alignment laser pulse, and the rotational
temperature of the molecules. Moreover, different probing schemes can influence the observed degree
of alignment. Here, the alignment laser pulses, at LCLS and CFEL, are not expected to cause this big
discrepancy. Generally, for a given laser peak intensity, the degree of alignment decreases with
decreasing adiabaticity. To account for the non-adiabaticity of the molecular alignment, the rise
time of the laser pulses and the peak intensities have to be examined. Both laser pulses had a
comparable peak intensities in the focus. The rise times at LCLS and CFEL were 70 and 100~ps,
respectively. Considering that the largest classical rotation period of the molecule is about 2~ns,
it is expected that the alignment dynamics during the rising edges of the laser pulses at both
experiments are similar. This is supported by the observed saturation of the degree of alignment for
a cold molecular beam, at CFEL, for the investigated peak intensities.

An analysis of the degree of alignment of (complex) molecules in dependence of the pulse duration
has not yet been performed, but from other experiments we have learned that the degree of alignment
is nearly independent of the pulse duration for a given pulse energy, over a wide range of
intermediate-regime durations, as long as ionization is avoided~\cite{Trippel:PRA89:051401R}.
Generally, for complex molecules it will be most practical to obtain the optimal pulse duration, for
maximum alignment, experimentally and corresponding experiments are forthcoming, especially in
preparation of future campaigns at FELs.

The different probing schemes, \ie, single photon x-ray probe at LCLS and IR strong field ionization
at CFEL, should both be reliable probing schemes and not cause the discrepancy in the measured
degree of alignment. This was confirmed by NIR strong field ionization experiments at LCLS to probe
the degree of alignment while no x-rays were available. Here, no significant changes in the degree
of alignment compared to the one obtained by the single photon x-ray probe were observed under
otherwise identical experimental conditions. Therefore, we conclude that the rotational temperature
of the molecular ensemble causes the significant difference in the measured degree of alignment, an
effect that was described in detail elsewhere~\cite{Kumarappan:JCP125:194309,
   Holmegaard:PRL102:023001}. The temperature of the pulsed valve was approximately \celsius{35}
higher at LCLS than at CFEL, which leads to a somewhat higher initial rotational temperature in the
supersonic expansion. In addition the stagnation pressure of the helium seed gas was lower at LCLS
than at CFEL, leading to a less efficient cooling in the expansion. Moreover, at LCLS the pulsed
valve was only 18~cm away from the interaction zone, which is likely too short a distance to get a
molecular beam without any collisions between the contained atoms and
molecules~\cite{Luria:JPCA115:7362}, \ie, the molecular beam could not have reached its terminal
temperature.

The achieved degree of alignment is slightly lower than the $\cost=0.89$ previously obtained at LCLS
for a molecule with a similar polarizability anisotropy~\cite{Kuepper:PRL112:083002}. This can be
attributed to the colder molecular beam achieved in that experiment and the fully adiabatic
alignment in the 10-ns-duration pulses provided by a 30~Hz Nd:YAG laser system. However, while the
Nd:YAG laser system needed to be user supplied and required a special setup with significant
technical challenges and safety implications, TSLs are generally available at FEL facilities for
pump-probe experiments and typically support the full XFEL repetition rates.

Furthermore, the elliptically polarized laser pulses were expected to 3D align the molecules. The 3D
alignment was not independently measured by ion-imaging at LCLS, but demonstrated by the experiments
at CFEL using the same alignment laser polarization ellipticity. However, due to the warmer molecular
beam at LCLS, the molecules are also expected to be less confined in 3D. The final analysis of the
x-ray diffraction data will provide more details on this.

The described experiment should be directly extensible to larger (bio)molecules, such as the
building blocks of life or small model peptides. These systems often have a polarizability
anisotropy comparable to, or even larger than, the current sample and scalar polarizabilities that
essentially scale with mass. The main obstacle for the molecular-frame investigation of such systems
at FELs is the generation of cold ensembles with sufficient density. Related resolution limits of
diffraction experiments had previously been discussed~\cite{Spence:ActaCrystA61:237}.

\section{Summary}
\label{sec:summary}
We have shown that it is possible to use the direct output from the regenerative amplifier of the
in-house TSL at the CXI beamline at LCLS to strongly align molecules at the full LCLS repetition
rate. We have measured a degree of alignment of \cost~=~0.85. We reasoned that the degree of
alignment was limited by the intrinsic temperature of the molecular beam, which was optimized for
increased diffraction signal, rather than by the alignment-laser parameters. We have presented
evidence for 3D-alignment of our molecular sample. This allows to directly observe bond length and
angles in the molecule in an x-ray diffraction experiment.

The demonstrated technique is advantageous to, for instance, Nd:YAG-laser-based alignment as it
allows to utilize the full repetition rate of current FELs. Experiments at those facilities are
expensive and by adapting this approach one makes optimal use of the limited beam time. Current FEL
facilities already have synchronized TSL systems making the implementation of our setup fast,
convenient, and, arguably, easy. Moreover, considering more complex experiments in a fs-laser-pump
x-ray-probe scheme, or \emph{vice versa}, it is possible to use different beams of the same optical
laser source as alignment, pump, and probe pulses. This reduces the amount of lasers to be
synchronized, the amount of possible sources of errors, and, therefore, makes the experiment overall
more stable. Our approach will also allow to utilize a very large fraction, if not all, of the
nearly 27\,000~pulses per second at the upcoming European XFEL or the one million pulses of LCLS-II,
solely depending on the availability of a synchronized high-repetition-rate millijoule-level
near-infrared laser system at these facilities. We point out that our method does not rely on the
presence of a well-defined molecular recoil-frame and, thus, has advantages over coincidence
techniques~\cite{Ullrich:RPP66:1463} for molecular-frame investigations of complex molecules, even
at high repetition rates. Furthermore, the alignment approach allows for multiple isolated molecules
in the interaction volume.

Providing stretched pulses of high-power amplified TSL systems, for instance, with pulse energies on
the order of \mbox{$10~\text{mJ}$} and durations of \mbox{$\sim\!1~\text{ns}$}, would allow for
investigations of strongly-aligned very large molecules, for instance, through coherent diffractive
imaging.

\section{Acknowledgments}
\label{sec:acknow}
Besides DESY, this work has been supported by the Helmholtz Virtual Institute ``Dynamic Pathways in
Multidimensional Landscapes'', the Helmholtz Association ``Initiative and Networking Fund'', the
excellence cluster ``The Hamburg Center for Ultrafast Imaging -- Structure, Dynamics and Control of
Matter at the Atomic Scale'' of the Deutsche Forschungsgemeinschaft (CUI, DFG-EXC1074), and the
European Research Council through the Consolidator Grant 614507-COMOTION. F.\,W.\ acknowledges a
Paul Ewald fellowship by the Volkswagen Stiftung. D.\,R.\ and A.\,R.\ acknowledge support from the
Office of Basic Energy Sciences, U.\,S.\ Department of Energy. Parts of this research were carried
out at the Linac Coherent Light Source (LCLS) at the SLAC National Accelerator Laboratory. 
Use of the Linac Coherent Light Source (LCLS), SLAC National Accelerator Laboratory, is
supported by the U.S. Department of Energy, Office of Science, Office of Basic
Energy Sciences under Contract No. DE-AC02-76SF00515.

\bibliography{string,cmi}
\bibliographystyle{iopart-num}

\end{document}